\documentclass[aps]{revtex4}
\usepackage{amsmath}
\usepackage{graphicx}
\usepackage{epsfig}
\usepackage{color}

\begin{document}

\title{Flipping-shuttle oscillations of bright one- and two-dimensional
solitons in spin-orbit-coupled Bose-Einstein condensates with Rabi mixing}
\author{Hidetsugu Sakaguchi}
\address{Department of Applied Science for
Electronics and Materials, Interdisciplinary Graduate School of Engineering
Sciences, Kyushu University, Kasuga, Fukuoka 816-8580, Japan}
\author{Boris A. Malomed}
\address{
Department of Physical Electronics, School of Electrical Engineering, Faculty of
Engineering, and Center for Light-Matter Interaction, Tel Aviv University, Tel Aviv 69978, Israel\\
ITMO University, St. Petersburg 197101, Russia}

\begin{abstract}
We analyze a possibility of macroscopic quantum effects in the form of
coupled structural oscillations and shuttle motion of bright two-component
spin-orbit-coupled striped (one-dimensional, 1D) and semi-vortex
(two-dimensional, 2D) matter-wave solitons, under the action of linear
mixing (Rabi coupling) between the components. In 1D, the intrinsic
oscillations manifest themselves as flippings between spatially even and odd
components of striped solitons, while in 2D the system features periodic
transitions between zero-vorticity and vortical components of semi-vortex
solitons. The consideration is performed by means of a combination of
analytical and numerical methods.
\end{abstract}

\maketitle

\section{Introduction}

Atomic Bose-Einstein condensates (BEC), in addition to exhibiting a great
deal of their own dynamical regimes \cite{BEC-book1,Rev1,Rev2}, have drawn a
lot of interest as testing grounds for the emulation of various effects from
condensed-matter physics \cite{emulator}, a prominent example \ provided by
the spin-orbit coupling (SOC). Although the true spin of bosonic atoms, such
as $^{87}\mathrm{Rb}$, used for the SOC emulation in BEC, is zero, the wave
function of the condensate may be composed as a mixture of two components
representing different hyperfine atomic states. The resulting \textit{\
pseudospin} $1/2$ makes it possible to map the spinor wave function of
electrons in solids into the two-component bosonic wave function of the
atomic BEC. Breakthrough experiments \cite{Nature,soc2} have demonstrated
the real possibility to simulate the SOC effect in the bosonic gas, in the
form of the linear interaction between the momentum and pseudospin of
coherent matter waves. Two fundamental types of the SOC, well known from
works on physics of semiconductors, which are represented by the Dresselhaus
\cite{Dresselhaus} and Rashba \cite{Rashba} Hamiltonians, as well as the
Zeeman-splitting effect \cite{Campbell}, may be simulated in the atomic BEC.
While the initial experiments on the SOC emulation realized effectively
one-dimensional (1D) settings \cite{NatureRev,Review}, the implementation of
the SOC in an effectively 2D geometry was reported too \cite{2D-experiment}.

The SOC being a linear effect by itself, its interplay with the intrinsic
nonlinearity of the BEC, which is usually induced, in the mean-field
approximation, by contact inter-atomic collisions or long-range
dipole-dipole interactions, produces various localized structures, such as
vortices \cite{vortex1,vortex2,vortex3,SL,vortex4}, monopoles \cite{monopole}%
, skyrmions \cite{skyrmions2,skyrmions1}, {and dark solitons \cite{DS2,DS3}.
The use of periodic potentials, induced by optical lattices, offers
additional possibilities -- in particular, the creation of gap solitons \cite%
{gap-sol1,gap-sol2,gap-sol3}}.

The conventional repulsive sign of inter-atomic forces can be switched to
attraction by means of the Feshbach resonance \cite{Cornish,Feshbach}, which
suggests possibilities for the creation of bright matter-wave solitons \cite%
{Rice,Paris,Cornish2,Cornish3}, in addition to the well-known dark ones \cite%
{DS1}. In particular, the modulational instability \cite{MI} and various
options for the making of effectively 1D bright solitons under the action of
SOC in attractive condensates have been theoretically analyzed in detail
\cite{sol1}-\cite{He}. A challenging possibility is to introduce 2D bright
solitons, which are always unstable against the critical collapse in the
usual BEC models based on the nonlinear Schr\"{o}dinger -- Gross-Pitaevskii
equations (NLSEs-GPEs) with attractive cubic terms \cite{Fibich}. As
demonstrated in Ref. \cite{Sakaguchi}, the SOC terms break the specific
scaling invariance of the GPE system in the 2D space, lift the related
degeneracy of the norm of the respective 2D solitons, and thus push the norm
below the threshold necessary for the onset of the critical collapse,
securing their stability. This unique possibility to stabilize bright
solitons in the free 2D space was further elaborated in Refs. \cite{Cardoso}-%
\cite{Sherman}. Furthermore, the same mechanism may create free-space
metastable solitons in the 3D geometry, although in that case the solitons
cannot realize the system's ground state \cite{Pu}.

In addition to the realizations of SOC in BEC, the similarity between the
GPEs for the binary condensate and the NLSE system modeling the
copropagation of orthogonal polarizations of light in twisted nonlinear
optical fibers~\cite{old,sol2} suggests to link the SOC to a broad range of
nonlinear effects in optics. This link has been recently extended to 2D
setting too \cite{optics,NJP}, making it possible to predict stable
spatiotemporal optical solitons in planar dual-core\ waveguides.
Manifestations of SOC\ are also known in other photonic settings \cite%
{Bliokh}. In particular, SOC can be directly realized in exciton-polariton
fields trapped in semiconductor microcavities \cite{cavity}. Taking into
regard nonlinearity in the latter setting makes it possible to predict 2D
trapped modes similar to the solitons found in the 2D model of the BEC with
SOC \cite{Dmitry}.

A common feature of 1D and 2D bright solitons supported by the attractive
nonlinearity in the two-component system coupled by the spin-orbit
interaction is the different shape of solitons in the cases when the
XPM/SPM\ ratio (the relative strength of the cross-attraction and
self-attraction), $\gamma $, takes values $\gamma \leq 1$ or $\gamma \geq 1$%
. In the former case, the 1D system produces stable \textit{striped solitons}
(see, e.g., Ref. \cite{we}), built as patterns featuring multiple density
peaks in the two components, with density maxima of one component coinciding
with minima of the other. Accordingly, the two components of the striped
solitons feature opposite spatial parities, one being even and the other
odd. In the case of $\gamma \geq 1$, stable 1D solitons feature a smooth
single-peak density profile, identical for both components. Similarly, the
2D system with $\gamma \leq 1$ supports \textit{semi-vortex} (SV)\textit{\ }%
solitons as stable modes, with isotropic components which carry,
respectively, vorticities $0$ and $1$, while stable solitons produced by the
same system with $\gamma \geq 1$ are \textit{mixed modes}, which combine
terms with zero and nonzero vorticities in each component \cite{Sakaguchi}.
Precisely at $\gamma =1$ (the Manakov's nonlinearity \cite{Manakov}),
solitons of both types stably coexist \cite{Sakaguchi,we}.

Because bright matter-wave solitons, predicted and observed in BEC, are
\textit{macroscopic quantum objects}, the consideration of the overall
dynamics of solitons in binary condensates under the action of SOC suggests
a possibility to observe \emph{macroscopic manifestations} of SOC. An
example is provided by recent work \cite{Beijing}, in which an artificial
magnetic field, induced\ by the SOC terms in the 1D system, drives
precession of the soliton's pseudospin, which, in turn, drives shuttle
motion of the 1D soliton as a whole. Prior to that, coupling of the
precession of the total pseudospin to the motion of a dark soliton in a
ring-shaped effectively one-dimensional SOC system was predicted in Ref.
\cite{Brand}.

The objective of the present work is to report another kind of macroscopic
dynamical effects featured by 1D and 2D solitons alike, under the combined
action of the SOC and Rabi coupling (RC). These effect exhibit periodic
\textit{flippings} between the two components of the condensate, coupled to
the shuttle motion of the soliton's center, with the same period. In the 1D
system, these are flippings between spatially even and odd components of
striped solitons, while in the 2D setting the vorticity is periodically
exchanged between two components of the SV soliton, if the motion of the
soliton is restricted in one direction by a quasi-1D confining potential.
The latter dynamical effect somewhat resembles periodic transfer of a single
vortex between two Rabi-coupled components of a 2D condensate, with the
repulsive nonlinearity acting in each component \cite{AFetter}, although in
that case the mode periodically exchanged between the components is not a\
bright soliton, but rather a vortex supported by the modulationally stable
background. As for the RC, it represents linear mixing between two hyperfine
atomic states (which constitute the two components), induced by a resonant
electromagnetic (GHz-frequency) wave coupling the atomic levels \cite{RC1}-%
\cite{RC5}.

The rest of the paper is organized as follows. The flipping-shuttle motion
of 1D solitons in considered, by means of analytical approximations and
systematic simulations, in Section II. The regime of periodic flippings
between the 2D\ SV soliton and its mirror-image counterpart, coupled to the
shuttle motion of the soliton as a whole in the direction which is not
restricted by the confining potential, is investigated, chiefly by means of
numerical simulations, in Section III. It is also shown that the use of an
isotropic trapping potential, instead of the quasi-1D one, leads to chaotic
dynamics, instead of regular flipping-shuttle motion. The paper is concluded
by Section IV.

\begin{figure}[t]
\begin{center}
\includegraphics[height=3.5cm]{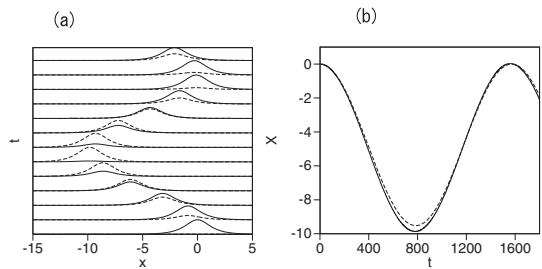}
\end{center}
\caption{(a) Snapshots of components $|\protect\phi _{+}(x)|$ and $|\protect%
\phi _{-}(x)|$ of the 1D soliton (shown by continuous and dashed lines,
respectively) at $t=150\times n$ ($n=1,2,\cdots ,12$). (b) The continuous
line depicts the motion of the soliton's center of mass, $X(t)$, for $%
\protect\lambda =0.02$, $d=0.002$, $\protect\gamma =1$, and total norm $N=2$%
, see Eq. (\protect\ref{N}). The dashed line is the analytical prediction
given by Eq. (\protect\ref{xi}).}
\label{fig1}
\end{figure}

\section{Flipping-shuttle dynamics of one-dimensional solitons}

\begin{figure}[t]
\begin{center}
\includegraphics[height=3.5cm]{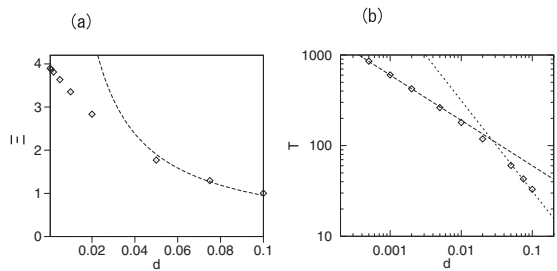}
\end{center}
\caption{(a) Half-amplitude $\Xi $ of the shuttle motion vs. the
Rabi-coupling strength, $d$, at $\protect\lambda =0.2$, $\protect\gamma =1$,
and $N=2$. The dashed line is the analytical prediction $\Xi =(10/21)(%
\protect\lambda /d)$ given by Eq. (\protect\ref{Xi}). (b) The period of the
flipping-shuttle oscillations vs. $d$ at $\protect\lambda =0.2$, $\protect%
\gamma =1$, and $N=2$. The dotted line is $\protect\pi /d$, which
corresponds to Eq. (\protect\ref{xi}), while the dashed line is $19/d^{1/2}$.
}
\label{fig2}
\end{figure}

We consider the GPE system with SOC terms of the Rashba type and RC, which
is written in a scaled form as
\begin{eqnarray}
i\frac{\partial \phi _{+}}{\partial t} &=&-\frac{1}{2}\frac{\partial
^{2}\phi _{+}}{\partial x^{2}}-(|\phi _{+}|^{2}+\gamma |\phi _{-}|^{2})\phi
_{+}+\lambda \frac{\partial \phi _{-}}{\partial x}-d\phi _{-},  \notag \\
&&  \label{1DSOC} \\
i\frac{\partial \phi _{-}}{\partial t} &=&-\frac{1}{2}\frac{\partial
^{2}\phi _{-}}{\partial x^{2}}-(|\phi _{-}|^{2}+\gamma |\phi _{+}|^{2})\phi
_{-}-\lambda \frac{\partial \phi _{+}}{\partial x}-d\phi _{+},  \notag
\end{eqnarray}%
where $\gamma $ is the above-mentioned relative strength of the
inter-component attraction, with respect to the self-attraction. Previous
works, which addressed this system in the absence of the RC ($d=0$), have
revealed striped bright solitons, composed of alternating segments occupied
by the two components (with opposite parities, even and odd), at $\gamma <1$%
, and smooth solitons, with $\left\vert \phi _{+}(x)\right\vert =\left\vert
\phi _{-}(x)\right\vert $, at $\gamma >1$ \cite{we}. In fact, scattering
lengths of interactions between atoms which represent different components
of the pseudo-spinor wave function are almost exactly equal \cite{Ho},
therefore we focus below, chiefly, on the case of $\gamma =1$, which
corresponds the Manakov's nonlinearity, in terms of optics models \cite%
{Manakov} (nevertheless, the case of $\gamma \neq 1$ is briefly considered
too, see Fig. \ref{fig5} below). If SOC is absent ($\lambda =0$), the 1D
Manakov's system is integrable, including the case when the RC terms are
present \cite{Tratnik}. In the latter case, it is easy to find an exact
bright-soliton solution with Rabi oscillations between the components:
\begin{equation}
\phi _{+}=\frac{A\cos (dt)e^{iA^{2}t/2}}{\cosh (Ax)},\;\phi _{-}=\frac{%
iA\sin (dt)e^{iA^{2}t/2}}{\cosh (Ax)}.  \label{fp}
\end{equation}%
The total norm and energy of the general 1D system (\ref{1DSOC}),\ which
includes the SOC and RC terms, are%
\begin{equation}
N=\int_{-\infty }^{+\infty }\left( \left\vert \phi _{+}(x)\right\vert
^{2}+\left\vert \phi _{-}(x)\right\vert ^{2}\right) dx,  \label{N}
\end{equation}%
\begin{equation}
E=\int_{-\infty }^{+\infty }\left\{ \frac{1}{2}\left( \frac{\partial \phi
_{+}}{\partial x}\right) ^{2}+\frac{1}{2}\left( \frac{\partial \phi _{-}}{%
\partial x}\right) ^{2}-\frac{1}{2}\left( |\phi _{+}|^{2}+|\phi
_{-}|^{2}\right) ^{2}+\frac{\lambda }{2}\left( \phi _{+}^{\ast }\frac{%
\partial \phi _{-}}{\partial x}-\phi _{-}^{\ast }\frac{\partial \phi _{+}}{%
\partial x}+\mathrm{c.c.}\right) \right\} dx,  \label{en}
\end{equation}%
where $\mathrm{c.c.}$\ stands for the complex-conjugate expression. When RC
is absent, $d=0$, and SOC is weak, i.e., $\lambda $ is small, an approximate
solution to Eq. (\ref{1DSOC}) with a large even component, $\phi _{+}$, and
a small odd one, $\phi _{\_}$ (these assumptions are suggested by the
presence of the weak SOC terms), may be sought for as
\begin{equation}
\phi _{+}=\frac{Ae^{iA^{2}t/2}}{\cosh (Ax)},\;\phi _{-}=\frac{B\sinh
(Ax)e^{iA^{2}t/2}}{\cosh ^{2}(Ax)},  \label{exact}
\end{equation}%
with $B^{2}\ll A^{2}$. The substitution of this ansatz in Eq.~(\ref{en})
yields
\begin{equation}
E=\frac{7}{15}AB^{2}-\frac{B^{4}}{35A}+\frac{4}{3}\lambda AB-\frac{1}{3}%
A^{3}.  \label{E}
\end{equation}%
For given $A$ and small $\lambda $, the corresponding small amplitude $B$ is
predicted by the variational equation $\partial E/\partial B=0$:
\begin{equation}
B=(-10/7)\lambda +O\left( \lambda ^{2}/A^{2}\right) .  \label{B}
\end{equation}

Proceeding to simulations of the full GPE system (\ref{1DSOC}), but, at
first, with small SOC and RC terms, Figs. \ref{fig1}(a) and (b) show,
respectively, snapshots of profiles of $|\phi _{+}(x)|$ and $|\phi _{-}(x)|$
at time moments $t=150\times n$ ($n=1,2,\cdots ,12$), and the corresponding
numerically found law of motion of the soliton's center-of-mass coordinate, $%
X(t)$, obtained for parameters $d=0.002$, $\lambda =0.02$, $\gamma =1$ and $%
N=2$. For these simulations, initial conditions, $\phi _{+}(x,t=0)$ and $%
\phi _{-}(x,t=0)$, were produced as stationary solutions of Eq.~(\ref{1DSOC}%
) with $d=0$ (but $\lambda \neq 0$), by dint of the imaginary-time-evolution
method. Due to the presence of the SOC terms, the two components have
opposite spatial parities: $\phi _{+}(-x,t=0)=\phi _{+}(x,t=0),~\phi
_{-}(-x,t=0)=-\phi _{-}(x,t=0)$. Thus, Fig. \ref{fig1} demonstrates that, at
$d\neq 0$, shuttle motion appears, coupled to flipping oscillations. For an
analytical consideration of this dynamical regime, we adopt an ansatz in the
form which is also suggested by exact solution (\ref{fp}) of the Manakov's
system, but, unlike the above ansatz (\ref{exact}), this time it combines
expressions modeled on solution (\ref{fp}) in both components:%
\begin{gather}
\left(
\begin{array}{c}
\phi _{+} \\
\phi _{-}%
\end{array}%
\right) =\exp \left[ i\frac{A^{2}}{2}t+i\frac{d\xi }{dt}\left( x-\xi \right) %
\right] \mathrm{sech}\left[ A(x-\xi )\right]   \notag \\
\times \left(
\begin{array}{c}
A\cos (dt)-iB\sin (dt)\tanh \left[ A(x-\xi )\right]  \\
iA\sin (dt)+B\cos (dt)\tanh \left[ A(x-\xi )\right]
\end{array}%
\right) ,  \label{ap2}
\end{gather}%
where $\xi (t)$ is the central position of the soliton. Then, applying the
variational approximation to this system leads, eventually, to the same
relation (\ref{B}) between the small and large amplitudes, $A$ and $B$.

\begin{figure}[tbp]
\begin{center}
\includegraphics[height=3.5cm]{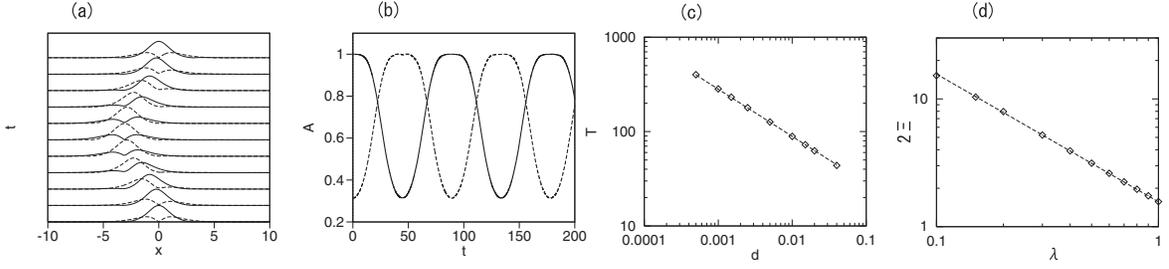}
\end{center}
\caption{(a) The evolution of components $|\protect\phi _{+}(x)|$ and $|%
\protect\phi _{-}(x)|$ (solid and dashed lines) produced by simulations of
the 1D system~(\protect\ref{1DSOC}) with $\protect\gamma =1$, $d=0.01$, $%
\protect\lambda =0.5$, and $N=2$. (b) Amplitudes of $|\protect\phi _{+}(x)|$
and $\left\vert \protect\phi _{-}(x)\right\vert $(solid and dashed curves,
respectively) as functions of time. (c) Period $T$ of the shuttle-flipping
oscillaltions vs. the Rabi-coupling coefficient, $d$, shown on the log-log
scale, for $N=2$ and $\protect\lambda =0.5$. The dashed line is $%
T=8.9d^{-1/2}$. (d) The amplitude of the shuttle motion, $2\Xi $, vs. the
SOC coefficient, $\protect\lambda $, on the log-log scale, for $N=2$ and $%
d=0.005$. The dashed line is $2\Xi =1.56\protect\lambda ^{-1}$.}
\label{fig3}
\end{figure}

Further, to address the motion of the soliton as a whole, it is relevant to
consider the total momentum of the system,
\begin{equation}
P=i\int_{-\infty }^{+\infty }\left( \frac{\partial \phi _{+}^{\ast }}{%
\partial x}\phi _{+}+\frac{\partial \phi _{-}^{\ast }}{\partial x}\phi
_{-}\right) dx.  \label{mo}
\end{equation}
Being a dynamical invariant of Eq. (\ref{1DSOC}), $P$\ keeps zero initial
value. On the other hand, ansatz (\ref{ap2}), if substituted in Eq.~(\ref{mo}%
), produces
\begin{equation}
P=-\frac{4}{3}AB\sin (dt)\cos (dt)+2\left( A+\frac{B^{2}}{3A}\right) \frac{%
d\xi }{dt}.  \label{P}
\end{equation}%
Finally, substituting $B$, as given by Eq. (\ref{B}) for sufficiently small $%
\lambda $, in the momentum-conservation condition $P=0$ following from Eq. (%
\ref{P}), leads to the prediction for the velocity of the moving soliton:
\begin{equation}
\frac{d\xi }{dt}=-\frac{20}{21}\lambda \sin (2dt).  \label{velocity}
\end{equation}%
A solution of this equation, satisfying the initial condition $\xi (0)=0$,
is
\begin{gather}
\xi (t)=-\Xi _{\mathrm{pert}}\left[ 1-\cos (2dt)\right] ,  \label{xi} \\
\Xi _{\mathrm{pert}}=\left( 10/21\right) \left( \lambda /d\right) .
\label{Xi}
\end{gather}

Direct simulations of Eq. (\ref{1DSOC}), performed at sufficiently small $%
\lambda $, produce the shuttle motion of the soliton's center-of-mass
coordinate
\begin{equation}
X=N^{-1}\int_{-\infty }^{+\infty }x(|\phi _{+}|^{2}+|\phi _{-}|^{2})dx,
\label{X}
\end{equation}%
which is very close to analytical prediction (\ref{xi}), as seen in Fig.~\ref%
{fig1}(b).

Figure \ref{fig2}(a) shows numerically evaluated half-amplitude of the
shuttle oscillations, $\Xi $ and its perturbative prediction (the dashed
line), given by Eq. (\ref{Xi}), for $\lambda =0.2$, $\gamma =1$, and $N=2$.
Figure \ref{fig2}(b) shows the numerically found period of the
flipping-shuttle oscillations, $T$ and its perturbative prediction, $\pi /d$
(the dotted line), for the same parameters. They demonstrate that the
predictions are fairly good unless the RC strength, $d$, becomes too small
(roughly, smaller than $\lambda /L$, where $L$ is a characteristic width of
the soliton), when it must be treated as a perturbation, see below [note
that, to derive Eq. (\ref{velocity}), the RC terms were taken into account
not perturbatively but as leading ones, while the SOC was treated as a
perturbation].

\begin{figure}[tbp]
\begin{center}
\includegraphics[height=3.5cm]{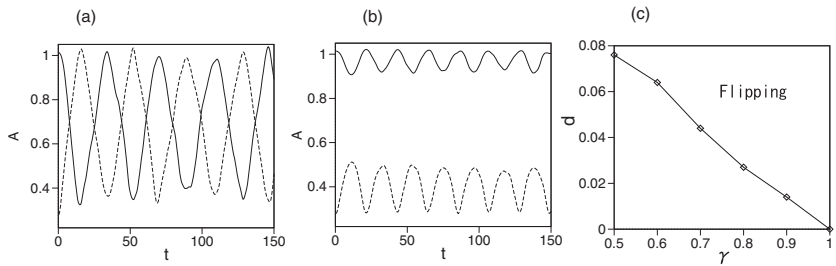}
\end{center}
\caption{(a) The evolution of amplitudes of components $|\protect\phi _{+}|$
and $\left\vert \protect\phi _{-}\right\vert $ (solid and dashed curves), at
$d=0.1$, $\protect\gamma =0.5$, $\protect\lambda =0.5$, and $N=2$. (b) The
same, but at $d=0.05$. In the former case, flippings take place, while in
the latter one flippings are suppressed. (c) The smallest value of $d$, at
which flippings commence (for $\protect\lambda =0.5$, $N=2$), vs. $\protect%
\gamma $.}
\label{fig4}
\end{figure}

For stronger SOC (larger $\lambda $), the shuttle-flipping dynamics was
studied by means of systematic simulations of Eq. (\ref{1DSOC}). Figure \ref%
{fig3}(a) shows snapshot profiles of $|\phi _{+}|$ and $|\phi _{-}|$ for $%
\lambda =0.5$ and $d=0.01$. In this case, flipping oscillations between even
and odd spatial components, accompanying the shuttle motion of the soliton
as a whole (with amplitude $2\Xi =3.08$) are clearly observed. Figure \ref%
{fig3}(b) illustrates this dynamical regime by displaying the evolution of
amplitudes of components $|\phi _{+}|$ and $|\phi _{-}|$. This dynamical
regime may be compared to a different one, which was reported, as mentioned
above, in Ref. \cite{Beijing}, which addressed the 1D system with SOC of the
mixed Rashba-Dresselhaus type, RC, and Zeeman detuning. In Ref. \cite%
{Beijing}, the variational approximation and direct simulations have
revealed shuttle oscillations of two-component solitons, both bright and
dark ones, coupled to the rotation of their pseudo-spin vectors around the
artificial magnetic field (the bright soliton suffered decay if the cubic
nonlinearity was not strong enough). Getting back to the present model, we
note that it also applies in fiber optics to the bimodal light propagation
in a nonlinear twisted fiber (the twist accounts for the effective RC
between two polarizations of light), if the phase-velocity and
group-velocity birefringence are taken into regard, emulating the Zeeman
splitting and SOC, respectively \cite{old}. In the fiber-optics model,
similar oscillations of the polarization of light, coupled to shuttle motion
of the soliton's center along the temporal coordinate, were predicted long
ago \cite{old}, and a related dynamical regime was proposed for the use in
\textit{rocking fiber-optics filters} \cite{Wabnitz}.

Figure \ref{fig3}(c) summarizes the numerical results by showing a
relationship between the period, $T$, of the flipping-shuttle oscillations
and small values of the RC coefficient, $d$, at $\lambda =0.5$. The figure
demonstrates scaling $T\sim d^{-1/2}$, which is clearly different from that
exhibited by the exact solution (\ref{fp}) of the Manakov's system, as well
as by the approximate solution (\ref{xi}) derived by means of the
perturbation theory for small $\lambda $ (while the RC terms were treated as
basic ones, rather than as a perturbation), $T_{0}=\pi /d$. Scaling $T\sim
d^{-1/2}$ can be explained by the fact that, if the RC represents a
perturbation, while the SOC terms are included in the main part of the
system (even if it is not easy to do that explicitly), the restoration
force, induced by the perturbation, scales as $d$, hence the frequency of
small oscillations, induced by this force, scales as $\sqrt{d}$. A global
picture of the $T(d)$ dependence is depicted in Fig. \ref{fig2}(b), showing
the crossover from $T=19/d^{1/2}$ at smaller $d$ to $T_{0}=\pi /d$ at larger
$d$.

Further, Fig. \ref{fig3}(d) displays the dependence of amplitude $2\Xi $ of
the shuttle motion on non-small values of the SOC coefficient, $\lambda $,
for fixed small $d=0.005$. The dependence suggests scaling $\Xi \sim \lambda
^{-1}$, which is strongly different from that in Eq. (\ref{Xi}), derived
above for small $\lambda $. This scaling can be readily explained in the
limit of large $\lambda $. Indeed, as shown in Refs. \cite{NJP} and \cite%
{Raymond}, for large $\lambda $ one may neglect, in the first approximation,
the kinetic-energy terms in Eq. (\ref{1DSOC}), which lends the system a
quasi-Dirac spectrum with a gap, $\omega ^{2}=d^{2}+\lambda ^{2}k^{2}$ ($%
\omega $ and $k$ are the frequency and wavenumber of small excitations),
keeping $\lambda ^{-1}$ as the \emph{single} spatial scale.

In the case of the Manakov's nonlinearity, considered above ($\gamma =1$),
flipping occurs at arbitrarily small values of the RC strength, $d$, which
is explained by the fact that this form of the nonlinearity supports
rotational invariance in the plane of the two components, $\left( \phi
_{+},\phi _{-}\right) $, thus facilitating their mutual conversion. However,
at $\gamma <1$ there is a barrier against the conversion, which prevents
flippings at small $d$. Figure \ref{fig4}(a) illustrates this effect,
showing that flippings take place at $d=0.1$ for $\gamma =0.5$, $\lambda =0.5
$, and $N=2$. On the other hand, in is seen in Fig. \ref{fig4}(b) that
flippings are suppressed at $d=0.05$ (the amplitude of $|\phi _{+}|$ always
remains larger that that of $|\phi _{-}|$). The evolution displayed in Figs. %
\ref{fig4}(a) and (b) shows some irregularity at $\gamma \neq 1$, due to the
fact that the evolution was initiated by the initial condition constructed
as the stationary solution of Eq.~(\ref{1DSOC}) with $d=0$, while the
simulations were performed with $d\neq 0$. Further, Fig. \ref{fig4}(c) shows
the critical (smallest) value of $d$ at which flippings commence. To explain
the nearly linear dependence between critical $d$ and $1-\gamma $, we recall
the above-mentioned argument, according to which the RC\ terms, if treated
as a perturbation, induce a force (torque) $\sim d$ driving the linear
conversion in the plane of $\left( \phi _{+},\phi _{-}\right) $. On the
other hand, the barrier blocking the rotation is proportional to $1-\gamma $
(the deviation from the Manakov's case, $\gamma =1$). The onset of flipping
is determined by the equilibrium between these factors, i.e., indeed, $d\sim
1-\gamma $.

At $\gamma >1$, the 1D bright smooth solitons with $\left\vert \phi
_{+}(x)\right\vert =\left\vert \phi _{-}(x)\right\vert $ have a lower energy
than the striped ones (which exist at $\gamma >1$ too), and the smooth
solitons do not exhibit the flipping dynamics. Nevertheless, simulations
performed at $d>1$ with the input in the form of the striped solitons
demonstrate that regular flipping-shuttle dynamics still occurs, as shown in
Fig. \ref{fig5}(a) and (b) for $\gamma =1.25$, $d=0.05$, $\lambda =0.2$, and
$N=2$. The increase of $\gamma $ from $1.25$ to $1.30$, and of $d$ from $0.05
$ to $0.1$ leads to chaotization of the flipping dynamics, as shown in Fig.~%
\ref{fig5}(c).
\begin{figure}[tbp]
\begin{center}
\includegraphics[height=3.5cm]{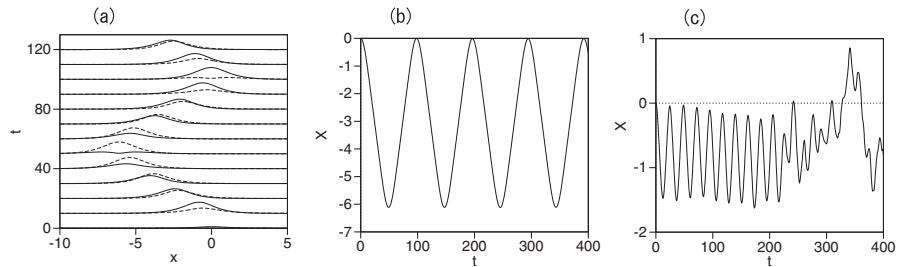}
\end{center}
\caption{(a) The evolution of components $|\protect\phi _{+}|$ and $%
\left\vert \protect\phi _{-}\right\vert $ (solid and dashed curves) in the
case of \ $\protect\gamma =1.25$; (b) the respective motion of the soliton's
center of mass. In this case, robust flipping-shuttle dynamics is observed
at $\protect\gamma >1$. Other parameters are $\protect\lambda =0.2$, $d=0.05$%
, and $N=2$. (c) Chaotic motion of the center of mass at $d=0.1$ for $%
\protect\gamma =1.3$ and $\protect\lambda =0.2$.}
\label{fig5}
\end{figure}
\begin{figure}[tbp]
\begin{center}
\includegraphics[height=3.5cm]{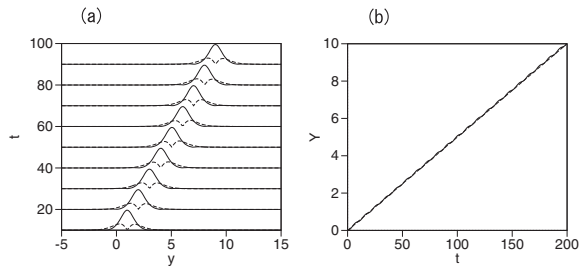}
\end{center}
\caption{(a) The evolution of cross-sections $x=0$ of two-dimensional
components $|\protect\phi _{+}(x,y)|$ and $|\protect\phi _{-}(x,y)|$ of the
semi-vortex soliton (solid and dashed lines, respectively), produced as a
numerical solution (semi-vortex) of Eq. (\protect\ref{2d}) at $d=0.05$, $%
\protect\gamma =1$, $\protect\lambda =1$, with total 2D norm $N=5$, see Eq. (%
\protect\ref{N2D}). (b) The evolution of coordinate $Y$ of the soliton's
center, defined as per Eq. (\protect\ref{XY}). The dashed line is $%
Y=0.05\cdot t$.}
\label{fig6}
\end{figure}
\begin{figure}[tbp]
\begin{center}
\includegraphics[height=3.5cm]{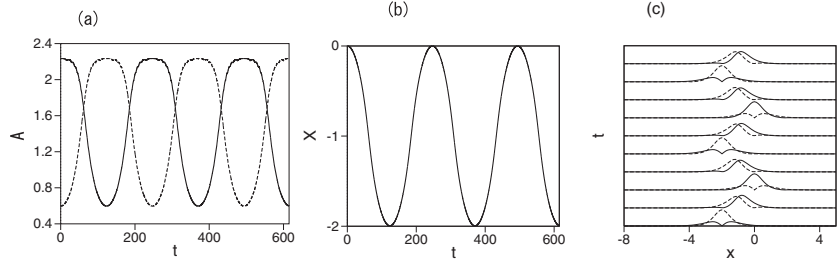}
\end{center}
\caption{(a) \ The evolution of amplitudes of components $|\protect\phi %
_{+}\left( x,y\right) |$ and $\left\vert \protect\phi _{-}\left( x,y\right)
\right\vert $ (solid and dashed lines, respectively) of the 2D semi-vortex,
obtained from simulations of Eq. (\protect\ref{2d}), at $\protect\gamma =1$,
$d=0.001$, $\protect\lambda =1$, $\Omega _{x}=0$, $\Omega _{y}=1$, and $N=5$%
. (b) The motion of the soliton's center in the $y$ direction, in the course
of the shuttle motion. (c) Cross-section profiles of components $|\protect%
\phi _{+}\left( x,y\right) |$ and $\left\vert \protect\phi _{-}\left(
x,y\right) \right\vert $, drawn along $y=0$ (solid and dashed lines), at $%
t=61.5\times n$ ($n=0,1,\cdots ,10$).}
\label{fig7}
\end{figure}
\begin{figure}[tbp]
\begin{center}
\includegraphics[height=3.5cm]{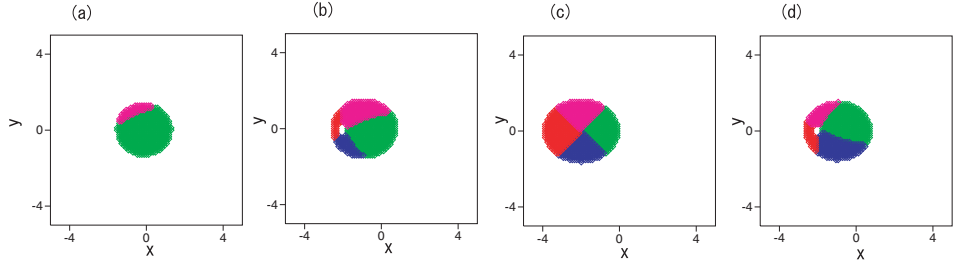}
\end{center}
\caption{Four snapshot profiles of $\protect\phi _{+}$, displayed at (a) $%
t=30.75$, (b) $t=61.5$, (c) $t=123$, and (d) $t=184.5$, illustrate the dynamics
of the flipping evolution of a 2D semi-vortex, presented in Fig. \protect\ref%
{fig7}. Different colors cover four regions, defined by $|\protect\phi %
_{+}|>0.1$,$\;\mathrm{Re}\left( \protect\phi _{+}\right) >0$,$\;\mathrm{Im}%
\left( \protect\phi _{+}\right) >0$ (green) ; $|\protect\phi _{+}|>0.1$,$\;\mathrm{Re}
\left( \protect\phi _{+}\right) <0$,$\;\mathrm{Im}\left( \protect\phi %
_{+}\right) >0$ (blue); $|\protect\phi _{+}|>0.1$,$\;\mathrm{Re}\left( \protect\phi
_{+}\right) <0$,$\;\mathrm{Im}\left( \protect\phi _{+}\right) <0$ (red); $|\protect%
\phi _{+}|>0.1$,$\;\mathrm{Re}\left( \protect\phi _{+}\right) >0$,$\;\mathrm{%
Im}\left( \protect\phi _{+}\right) <0$ (purple). The junction point of the four
colors (sometimes seen as a while dot) is the pivot of the vortex, which
enters the zero-vorticity component from outside through the edge, attains the central
position, and then moves backwards, exiting the component through the same edge.}
\label{fig7b}
\end{figure}
\begin{figure}[tbp]
\begin{center}
\includegraphics[height=3.5cm]{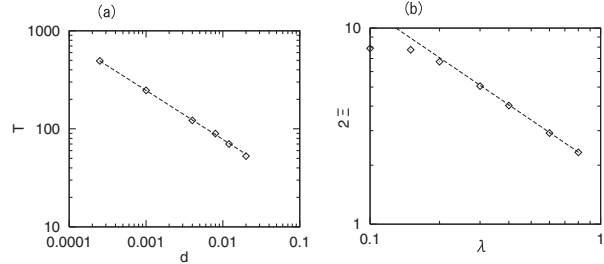}
\end{center}
\caption{(a) The period of the flipping-shuttle motion of a robust 2D
semivortex vs. $d$ at $\protect\lambda =1$. (b) The amplitude $2\Xi $ vs. $%
\protect\lambda $ of the shuttle motion at $d=0.01$. Other parameters are
the same as in Fig. \protect\ref{fig7}.}
\label{fig8}
\end{figure}
\begin{figure}[tbp]
\begin{center}
\includegraphics[height=3.5cm]{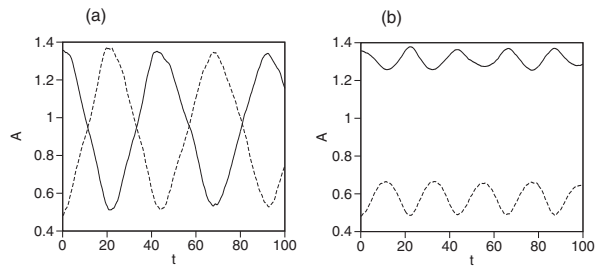}
\end{center}
\caption{The evolution of amplitudes of components $|\protect\phi _{+}\left(
x,y\right) |$ and $\left\vert \protect\phi _{-}\left( x,y\right) \right\vert
$(solid and dashed lines, respectively) of the 2D semi-vortex soliton at $%
d=0.1$ (a) and $d=0.05$ (b). Other parameters are $\protect\gamma =0$, $%
\protect\lambda =1$, $\Omega _{x}=0$, $\Omega _{y}=1$, and $N=4$, in both
cases.}
\label{fig9}
\end{figure}
\begin{figure}[t]
\begin{center}
\includegraphics[height=3.5cm]{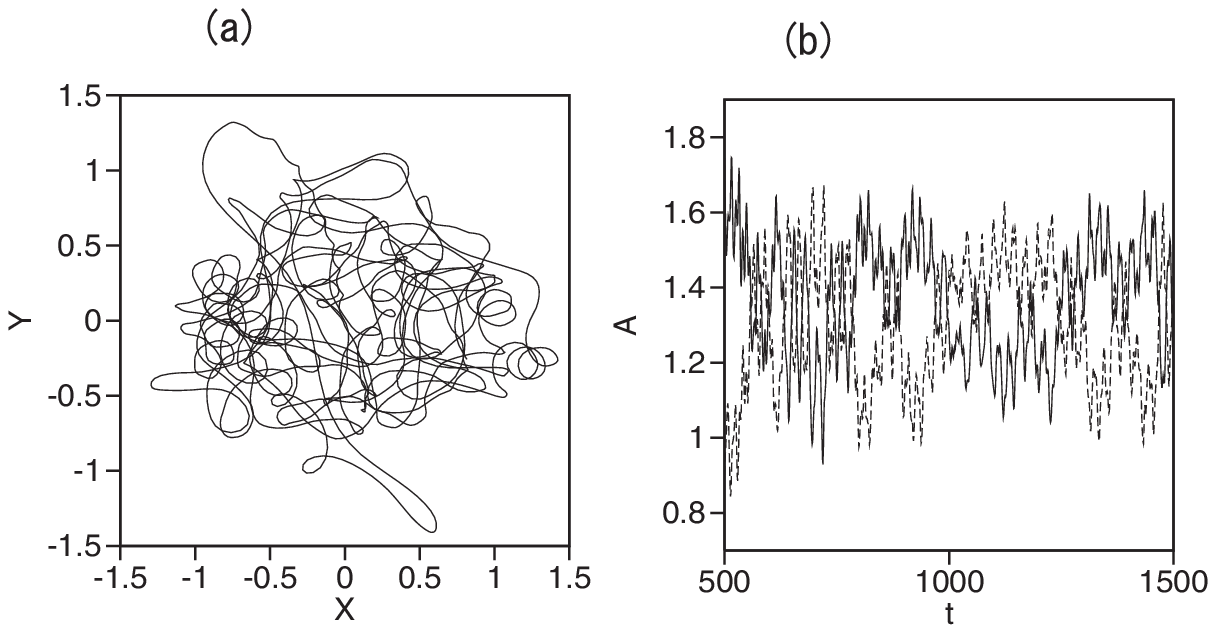}
\end{center}
\caption{(a) The trajectory of chaotic motion of the center of mass of the
semi-vortex soliton under the action of a shallow isotropic HO trapping
potential, with $\Omega _{x}=\Omega _{y}=0.01$, other parameters being $%
d=0.02$, $\protect\gamma =1$, $\protect\lambda =1$. (b) The evolution of
amplitudes of components $|\protect\phi _{+}\left( x,y\right) |$ and $%
\left\vert \protect\phi _{-}\left( x,y\right) \right\vert $ (solid and
dashed curves, respectively) in the same case. }
\label{fig10}
\end{figure}

\section{Flipping-shuttle dynamics of two-dimensional SV (semi-vortex)
solitons}

The 2D GPE system, which includes the SOC of the Rashba type (again, with
coefficient $\lambda $) and the RC terms in 2D, along with an
harmonic-oscillator (HO)\ trapping potential (generally, an anisotropic one,
with confining frequencies $\Omega _{x}\neq \Omega _{y}$), is written as a
straightforward extension of the model considered in Ref. \cite{Sakaguchi}:%
\begin{eqnarray}
i\frac{\partial \phi _{+}}{\partial t} &=&-\frac{1}{2}\nabla ^{2}\phi
_{+}-(|\phi _{+}|^{2}+\gamma |\phi _{-}|^{2})\phi _{+}+\lambda \left( \frac{%
\partial \phi _{-}}{\partial x}-i\frac{\partial \phi _{-}}{\partial y}%
\right) +\frac{1}{2}\left( \Omega _{x}^{2}x^{2}+\Omega _{y}^{2}y^{2}\right)
\phi _{+}-d\phi _{-},  \notag \\
i\frac{\partial \phi _{-}}{\partial t} &=&-\frac{1}{2}\nabla ^{2}\phi
_{-}-(|\phi _{-}|^{2}+\gamma |\phi _{+}|^{2})\phi _{-}-\lambda \left( \frac{%
\partial \phi _{+}}{\partial x}+i\frac{\partial \phi _{+}}{\partial y}%
\right) +\frac{1}{2}\left( \Omega _{x}^{2}x^{2}+\Omega _{y}^{2}y^{2}\right)
\phi _{-}-d\phi _{+}.  \label{2d}
\end{eqnarray}%
At $d=0$ and $\gamma \leq 1$, Eqs. (\ref{2d}) in free space (with $\Omega
_{x,y}=0$) give rise to stable bright solitons in the form of the SVs,
composed of an isotropic wave field with zero vorticity in one component,
and a solitary vortex in the other. Loosely speaking, the SVs may be
considered as the 2D generalization of the 1D striped solitons (in
particular, the difference of even and odd parities of the two components of
the 1D solitons resembles the difference of the zero and nonzero vorticities
of the SV's components). Note that, although coordinates $x$ and $y$ in the
free-space version of Eq. (\ref{2d}) appear differently, the equations are
invariant with respect to a change of the notation which readily swaps $x$
and $y$: $\tilde{\phi}_{+}\equiv \phi _{+}$, $\tilde{\phi}_{-}\equiv i\phi
_{-}$, $\tilde{x}\equiv -y$, $\tilde{y}\equiv x$. The total norm of the 2D
soliton is%
\begin{equation}
N=\int \int \left[ |\phi _{+}\left( x,y\right) |^{2}+\left\vert \phi
_{-}\left( x,y\right) \right\vert ^{2}\right] dxdy.  \label{N2D}
\end{equation}

Similar to the 1D system, we here focus on the Manakov's nonlinearity, with $%
\gamma =1$, which is quite close to the physically relevant situation, as
mentioned above. First of all, in the free space ($\Omega _{x,y}=0$),
results reported in Ref. \cite{Sakaguchi} actually demonstrate that, under
the action of the RC terms with strength $d$, the SV moves at a constant
velocity,
\begin{equation}
v_{y}=d/\lambda .  \label{vy}
\end{equation}%
Indeed, the transformation of Eq. (\ref{2d}) with $d=0$ and $\Omega _{x,y}=0$
into a reference frame moving in the $y$ direction with velocity $v_{y}$,
which is carried out by means of the substitution,%
\begin{equation}
\phi _{\pm }(x,y;t)=\tilde{\phi}_{\pm }\left( x,y-v_{y}t;t\right) \exp \left[
iv_{y}y-\left( i/2\right) v_{y}^{2}t\right] ,  \label{tilde}
\end{equation}%
generates effective RC terms with $\tilde{d}=-\lambda v_{y}$ , which
compensate the RC terms in Eq. (\ref{1DSOC}), thus making the existence of
the solitons moving at velocity (\ref{vy}) obvious. In exact accordance with
this, simulations of Eq. (\ref{2d}) with $\Omega _{x.y}=0$ produce stable
SVs moving at a constant velocity in the $y$ direction, as shown in Fig. \ref%
{fig6}. The initial condition is taken not as Eq.~(\ref{tilde}) at $t=0$,
but as the stationary SV state of Eq.~(\ref{2d}) with $\Omega _{x,y}=0$ and
without the RC term. The center of mass of the 2D solitons is defined as
\begin{equation}
\left\{ X,Y\right\} =N^{-1}\int \int \left\{ x,y\right\} (|\phi
_{+}|^{2}+|\phi _{-}|^{2})dxdy  \label{XY}
\end{equation}%
[cf. Eq. (\ref{X}) in the 1D model], where $N$ is the 2D norm defined as per
Eq. (\ref{N2D}). The numerically found velocity, $v_{y}\equiv dY/dt=0.05$,
corresponding to the situation displayed in Fig. \ref{fig6}, precisely
agrees with the value given by Eq. (\ref{vy}). Note that the same argument
is not valid for the 1D system (\ref{1DSOC}), as the application of the
transformation similar to that defined by Eq. (\ref{tilde}) to system (\ref%
{1DSOC}) does not generate the RC terms.

Similar to the flipping oscillations between the even and odd components of
1D stripe solitons, in 2D\ the RC may cause periodic flippings between the
SV with vorticities $\left( 0,+1\right) $ in its components $\left( \phi
_{+},\phi _{-}\right) $, and its mirror-image counterpart, with the
vorticity set $\left( -1,0\right) $, provided that free motion in the $y$
direction (see Fig. \ref{fig6}) is arrested by the confining HO potential in
Eq. (\ref{2d}) with $\Omega _{y}$ large enough, while $\Omega _{x}=0$ may be
set, the confinement in the $x$ direction being unnecessary, as suggested by
the results for the 1D system presented above. The initial condition for the
direct numerical simulation of Eq.~(\ref{2d}) is the stationary ground state
of Eq.~(\ref{2d}) with $d=0$, which includes the HO potential with $\Omega
_{y}=1$ . A typical example of robust periodic flippings, coupled to the
shuttle motion of the SV as a whole, is presented in Fig. \ref{fig7}.
Further, Fig. \ref{fig7b} shows a cycle of the transformations of the
fundamental (zero-vorticity) component into a vortex and back. The vortex
enters the fundamental component ($\phi _{+}$) from the edge at the moment
of time close to $t=61.5$, attains the central position at $t=123$, and then
moves backwards, leaving the pattern through the same edge from
which it has entered at $t=184.5$, when $\phi _{+}$ recurs to the
zero-vorticity shape. The shuttle motion coupled to the intrinsic flippings
is synchronized with them: the soliton reaches the leftmost position of
$X\simeq 2$ at $t=123$, and returns to the center, $X=0$, at $t=246$.

Figure \ref{fig8}(a) shows, on the log-log scale, the dependence of the
evolution period on the RC strength, $T(d)$, for $\lambda =1$ and $\gamma =1$%
, which is found to be $T\sim 1/\sqrt{d}$. This dependence is essentially
the same as in the 1D system, cf. \ref{fig2} (c), and the same qualitative
explanation for it, which was presented above for the 1D case, applies in
the present case as well. Figure \ref{fig8}(b) shows, on the log-log scale,
the respective dependence $2\Xi (\lambda )$ at $d=0.01$ and $\gamma =1$. The
dashed line is a line of $2\Xi =1.95/\lambda ^{0.8}$.

Similar to the 1D system considered above, the analysis of the 2D model in
the ``under-Manakov" case, with $\gamma <1$, demonstrates that the flipping
oscillations do not occur at too small values of the RC strength, $d$. In
particular, Fig. \ref{fig9}(a) shows that stable flippings take place at $%
d=0.1$, while other parameters are fixed as $\lambda =1$, $\gamma =0$, and $%
N=4$, but the flipping regime does not occur in Fig. \ref{fig9}(b) at $d$
reduced to $0.05$, when the amplitudes of the two components oscillate
without crossing zero. In this case, the critical value at which the
flipping regime sets in is $d_{c}\approx 0.08$, cf. Fig. \ref{fig4}(c) in
the 1D case.

Lastly, it is relevant to stress that the presence of the anisotropic HO
trapping potential, which acts only along the $y$ direction in the case
corresponding to Figs. \ref{fig7} and \ref{fig8}, is essential for
supporting the robust flipping-shuttle dynamical regime for the SVs in the
2D geometry. If an isotropic HO trapping potential is used, with $\Omega
_{x}=\Omega _{y}$, the evolution of the SV becomes chaotic under the action
of the RC, and regular shuttle motion is not observed, as shown in Fig. \ref%
{fig10}. A possible explanation to this may be the mismatch between the
isotropic shape of the trapping potential and anisotropic structure of the
SOC operator in Eq. (\ref{2d}). In a detailed form, this situation may be 
analyzed in a finite-mode approximation, expanding the two components of
the wave function over a truncated set of eigenstates of the isotropic HO
Hamiltonian, and, accordingly, replacing the coupled GPEs by a system of
 ordinary differential equations for the evolution of amplitudes of the 
truncated expansion (cf. Ref. \cite{Driben}), but detailed analysis of this 
approach is beyond the scope of the present work.

\section{Conclusion}

The objective of this work is to expand the variety of macroscopic quantum
effects produced by coherent evolution of matter waves in BEC. To this end,
we have considered the dynamics of 1D and 2D solitons in the binary SOC
(spin-orbit-coupled) system with intrinsic self- and cross-attractive
interactions, under the action of the linear RC\ (Rabi coupling). The latter
ingredient of the system can be readily induced by a resonant GHz-frequency
electromagnetic wave mixing different atomic states representing the two
components of the binary condensate. The RC gives rise to periodic flippings
between the spatially even and odd components of 1D striped solitons, and
between the zero-vorticity and vortical components of the stable 2D\
semi-vortices, in the presence of a quasi-1D confining potential. In both
cases, the intrinsic oscillations of the internal structure of the soliton
are coupled to the periodic shuttle motion of the soliton as a whole. These
results predict a possibility to observe new macroscopic manifestations of
the SOC in the matter-wave dynamics.

As an extension of the present analysis, it may be relevant to consider
interactions, including collisions, between 1D and 2D solitons performing
the flipping-shuttle oscillations. A challenging possibility is to extend
the consideration to the full 3D setting.

\section*{Acknowledgments}

The work of B.A.M. was supported, in part, by grant No. 2015616 from the
joint program in physics between the NSF and Binational (US-Israel) Science
Foundation, and by grant No. 1286/17 from the Israel Science Foundation.

\end{document}